\documentclass[3p,twocolumn]{elsarticle}

\usepackage{graphicx}
\usepackage{multirow}
\usepackage{mathrsfs}
\usepackage{bm}
\usepackage{amsmath}
\usepackage{amssymb}
\usepackage{color}
\usepackage{longtable}
\newcommand{\red}[1]{{#1}}

\begin{document}

\title{From deformed Hartree-Fock to the nucleon-pair approximation \tnoteref{t1}}

\author{G. J. Fu}
\ead{gjfu@tongji.edu.cn}
\address{School of Physics Science and Engineering, Tongji University,
1239 Siping Rd., Shanghai 200092, China}
\author{Calvin W. Johnson}
\ead{cjohnson@sdsu.edu}
\address{Department of Physics, San Diego State University,
5500 Campanile Drive, San Diego, CA 92182-1233}

\tnotetext[t1]{This article is registered under preprint arXiv:1909.08785}

\begin{abstract}
The nucleon-pair approximation (NPA) can be a compact alternative to full configuration-interaction (FCI) diagonalization in nuclear shell-model spaces, but selecting good pairs is a long-standing problem.
While seniority-based pairs  work well for near-spherical nuclides, they do not work well for deformed nuclides with strong rotational bands. We propose an alternate approach.
We show how one can write any Slater determinant for an even number of particles as a general pair condensate, from which one can project out pairs of good angular momentum. We implement this by generating unconstrained Hartree-Fock states in a shell model basis and extracting $S$, $D$, and $G$ pairs. The subsequent NPA calculations yield good agreement with FCI  results using the same effective interactions.
\end{abstract}



\maketitle

\section{Background and motivation}

The nuclear shell model  is a powerful framework for nuclear structure theory.  But full configuration-interaction (FCI), that is, diagonalization of a Hamiltonian using all configurations in a given single-particle space,
leads to exponentially exploding basis dimensions.
Hence, the hunt for efficient truncation schemes is a key challenge.
The nucleon-pair approximation (NPA) \cite{NPA1,NPA2,NPA3},  based on the pair truncation of the shell model 
configuration space, is one appealing approach.
The building blocks of the NPA are collective/noncollective fermion pairs with good angular momentum, such as $SD$ pairs (collective pairs with angular momentum zero and two).
The NPA is flexible enough to contain other well-known methods.
For example, if the model space contains only the collective $S$ pair, the NPA is exactly the generalized seniority scheme \cite{generalizedseniority,PhysRevC.85.034324,JPG.42.115105};
if all noncollective spin-zero pairs are considered, one obtains seniority truncation (exact pairing) of the shell model \cite{ca05};
if all possible fermion pairs are considered, the NPA configuration space is equivalent to FCI; and finally
if the Pauli principle is neglected, $SD$ pairs reduce to $sd$ bosons, the building blocks of the successful interacting boson model \cite{IBM,iachello1987interacting}.
The NPA has been widely applied to describe low-lying states of nearly-spherical nuclei in mass regions with $A \sim 80$, 100, 130, 210 (see Ref. ~\cite{NPAreview} for a recent review).
The competition between isovector and isoscalar pairing in  $N=Z$ nuclei has been investigated in the NPA with isospin symmetry \cite{NPAisospin,PhysRevC.91.054322}.
Finally, the configuration mixing of many major-shell orbits can be treated in the NPA with particle-hole excitations \cite{NPAph}.

The NPA has  proven to be a compact truncation, but selecting good pairs remains a long-standing problem.
In early applications of the NPA, which rested primarily upon  $SD$-pair truncation,  the structure coefficients of the collective $S$ pair  were found by solving the BCS equation, and the collective $D$ pair was obtained by the commutation between the quadrupole operator $\widehat{Q}$ and the  $S$ pair-creation operator, $\widehat{S}^{\dagger}$, i.e., $\widehat{D}^{\dagger} = [\widehat{Q},\widehat{S}^{\dagger}]$.
In recent years it has been shown that NPA calculations  can be improved if pair-structure coefficients are determined by the generalized seniority scheme, namely, the $S$ pair is chosen so that the expectation value of Hamiltonian in the $S$-pair condensate,
\begin{eqnarray}
\frac{ \langle (S_{\tau})^N | \hat{H} | (S_{\tau})^N \rangle   }{\langle (S_{\tau})^N | (S_{\tau})^N \rangle}, \quad  \text{with}~ \tau=\pi ~\text{or}~ \nu,
\end{eqnarray}
is minimized, and non-$S$ pairs  obtained by diagonalizing the Hamiltonian matrix in the space spanned by the generalized-seniority-two (i.e., one-broken-pair) states \cite{PhysRevC.79.054315}.
While seniority-based pairs provide a good descriptions of collective states in semimagic nuclei and vibrational nuclei \cite{PhysRevC.82.034303,PhysRevC.}, they do not work well for rotational bands in deformed nuclei.
For example, under the quadrupole-quadrupole interaction, the moment of inertia and the $E2$ transitions of the system with 6 valence protons and 6 valence neutrons in the $pf$ and $sdg$ shells calculated by the $SD$-pair approximation are much smaller than those obtained by the FCI \cite{PhysRevC.62.014316}.


In this Letter we  propose an alternate approach for even-even nuclides. 
We generate 
{ an unconstrained} Hartree-Fock (HF) state in a shell model basis, and then represent the Slater determinant as a pair condensate, from which we project out pairs of good angular momentum.
We find good agreement between the subsequent NPA calculations  and FCI diagonalization.
This is the first time   NPA calculations with realistic shell model 
interactions  have successfully reproduced rotational bands.

\section{Methods}

We start with unconstrained HF calculations in a shell model basis, that is, our HF states have arbitrary shape and orientation
(and even parity mixing if the space contains single-particle orbits of both parities)
without enforcing additional symmetries such as axial or time-reversal symmetry,   
using a previously developed code \cite{SHERPA}.
In general, the HF states  have nonzero expectation values of the quadrupole tensor, and for simplicity we call them deformed HF.
We use Greek letters $\alpha$, $\beta,\ldots$ to {label} the {original} single-particle states with quantum numbers, {including good angular momentum}, $n$, $l$, $j$, $m$, with fermion creation operator in this basis written as  $\hat{a}_{\alpha}^{\dagger}$.
{
The deformed single-particle states from our HF calculations, labeled by Latin letters $a$, $b,\ldots$ with creation operator $\hat{c}_{a}^{\dagger}$,
are a  transformation of the original single-particle states:
}
\begin{eqnarray}\label{eq2}
\hat{c}_{a}^{\dagger} = \sum_{\alpha} U_{a \alpha } \hat{a}_{\alpha}^{\dagger}.
\end{eqnarray}
{The columns of $U$ form orthonormal vectors, and so form part of a unitary transformation.}

A Slater determinant for an even number 
 of valence protons or neutrons can be written as a pair condensate:
\begin{align}\label{HF}
& \prod_{a=1}^{2N} \hat{c}_{a}^{\dagger} |0\rangle = (N!)^{-1} \left(\hat{ c}_{1}^{\dagger}\hat{c}_{2}^{\dagger} + \cdots + \hat{c}_{2N-1}^{\dagger}\hat{c}_{2N}^{\dagger} \right)^{N} |0\rangle   \nonumber \\
& \qquad \qquad = (N!)^{-1}\left(  \sum_{  ab} g_{ ab} ~ \hat{c}^{\dagger}_{a} \hat{c}^{\dagger}_{b} \right)^{N} |0\rangle
\end{align}
The r.h.s. of Eq. (\ref{HF}) is a pair condensate,
where $g_{12} = g_{34} = \ldots = g_{(2N-1)(2N)} =\pm 1 $, and other $g_{ij} =0$.
The ordering of 1,2,3,4... is arbitrary, as is the phase $\pm 1$ in front of each noncollective pair.
In general, for even-even nuclei the HF single-particle states have degenerate time-reversed partners, and for simplicity  we order by single-particle energy.

{
Using standard techniques \cite{ring2004nuclear} one can project out pairs of good angular momentum
from the deformed HF pair
\begin{equation}
\hat{\cal C}^\dagger = \sum_{ab} g_{ab} \hat{c}^\dagger_a \hat{c}^\dagger_b
=  \frac{1}{2} \sum_{\alpha \beta} C_{\alpha \beta} \hat{a}^\dagger_\alpha\hat{a}^\dagger_\beta
\end{equation}
where we've introduced the antisymmetric matrix
\begin{equation}
C_{\alpha \beta} = 
 \sum_{ab} g_{ab}
\left ( U_{a \alpha} U_{b \beta} - U_{b \alpha} U_{a \beta}
\right )
\end{equation}
To facilitate our development, we separate out the $j_z$ quantum numbers (and order $\alpha, \beta$ 
without loss of generality), writing
\begin{equation}
\hat{\cal C}^\dagger = \sum_{\alpha  \leq \beta; k_\alpha  k_\beta} C_{\alpha k_\alpha, \beta   k_\beta} \hat{a}^\dagger_{\alpha k_\alpha}
\hat{a}^\dagger_{\beta k_\beta},
\end{equation}
where $k_\alpha, k_\beta$ are the $z$-projections of angular momentum in the ``intrinsic'' state, and $\alpha, \beta$ now contain all other
quantum numbers.

 Now we apply the rotation operator $\hat{R}(\Omega)$, where $\Omega$ represents the Euler angles. For details see
 \cite{ring2004nuclear,edmonds1996angular}, but all
 we need are the Wigner $D$-matrices:  any state with good angular momentum $| J, K \rangle$ gets transformed under rotation as 
 \begin{equation}
 \hat{R}(\Omega) | J, K \rangle = \sum_M {\cal D}^{(J)}_{M,K} (\Omega) | J, M \rangle.
 \end{equation}
Applying this \red{to the pair creation operator},
\begin{eqnarray}
\hat{R}(\Omega) \hat{\cal C}^\dagger \red{\hat{R}^{-1}(\Omega) }
= \sum_ {\alpha \leq \beta;  k_\alpha k_\beta} C_{\alpha k_\alpha, \beta k_\beta}  \nonumber \\
\times \sum_{m_\alpha m_\beta}
{\cal D}^{(j_\alpha)}_{m_\alpha k_\alpha} (\Omega)
 \hat{a}^\dagger_{\alpha m_\alpha}
 {\cal D}^{(j_\beta)}_{m_\beta k_\beta} (\Omega)
\hat{a}^\dagger_{\beta m_\beta}.
\end{eqnarray}
But using \cite{edmonds1996angular}
\begin{eqnarray}
{\cal D}^{(j_\alpha)}_{m_\alpha k_\alpha} (\Omega)
 {\cal D}^{(j_\beta)}_{m_\beta k_\beta} (\Omega)
 = \sum_{J^\prime , \mu, \mu^\prime} {\cal D}^{(J^\prime)}_{\mu \mu^\prime} (\Omega)
  \nonumber \\
\times  (j_\alpha m_\alpha , j_\beta m_\beta | J^\prime \mu )
 (j_\alpha k_\alpha , j_\beta k_\beta | J^\prime \mu^\prime )
\end{eqnarray}
we get
\begin{eqnarray}
\hat{R}(\Omega) \hat{\cal C}^\dagger  \red{\hat{R}^{-1}(\Omega) }
= \sum_{\alpha \leq \beta; k_\alpha  k_\beta} C_{\alpha k_\alpha, \beta k_\beta}
\sum_{J^\prime, \mu, \mu^\prime}
\nonumber \\ \times
 {\cal D}^{(J^\prime)}_{\mu \mu^\prime} (\Omega)
 (j_\alpha k_\alpha , j_\beta k_\beta | J^\prime \mu^\prime )
\left [ \hat{a}^\dagger_\alpha \otimes \hat{a}^\dagger_\beta \right ]_{J^\prime, \mu}.
\end{eqnarray}
Now use the orthogonality of the $D$-matrices \cite{edmonds1996angular}
to project out a pair 
\begin{eqnarray}
\hat{B}^\dagger_{J,MK} =
\frac{2J+1}{8\pi^2}
\int d \Omega \, {\cal D}^{(J) *}_{M,K}(\Omega) \hat{R}(\Omega)\,  \hat{\cal C}^\dagger   \red{\hat{R}^{-1}(\Omega)}\nonumber  \\
=\sum_{\alpha \leq \beta; k_\alpha k_\beta} C_{\alpha k_\alpha, \beta k_\beta}
 (j_\alpha k_\alpha , j_\beta k_\beta | J K ) 
 \left [ \hat{a}^\dagger_\alpha \otimes \hat{a}^\dagger_\beta \right ]_{J M}  \nonumber 
\end{eqnarray}
\begin{equation}
 =  \sum_{\alpha \leq \beta}  y_{J,K}(\alpha \beta)  \left [ \hat{a}^\dagger_\alpha \otimes \hat{a}^\dagger_\beta \right ]_{J M}
\end{equation}
where the structure coefficients are defined by
\begin{equation}
y_{J,K}(\alpha \beta) =  \sum_{k_\alpha k_\beta} C_{\alpha k_\alpha, \beta k_\beta}
\frac{  (j_\alpha k_\alpha , j_\beta k_\beta | J K )}{1+\delta_{\alpha \beta}} .
\end{equation}

Of course, we want physically unique pairs, and not merely rotated copies, as well as 
results independent of the arbitrary orientation of the initial HF states.
Hence we compute the norm matrix,
\begin{eqnarray}
N^{(JM)}_{K^\prime K} =
\langle 0 | \hat{B}_{J,M K^\prime}
\hat{B}^\dagger_{J,MK} | 0 \rangle \nonumber \\
=   \sum_{\alpha \leq \beta}  (1+\delta_{\alpha \beta}) y^*_{J,K^\prime}(\alpha \beta) y_{J, K}(\alpha \beta) ,
\end{eqnarray}
where we've used \cite{bg77}
\begin{eqnarray}
\langle 0 |   \left [ \hat{a}^\dagger_\alpha \otimes \hat{a}^\dagger_\beta \right ]^\dagger_{J^\prime M^\prime}  
\left [ \hat{a}^\dagger_{\alpha} \otimes \hat{a}^\dagger_{\beta} \right ]_{J M}
|0 \rangle  \nonumber \\
= \delta_{J^\prime J} \delta_{M^\prime M} (1+ (-)^J \delta_{\alpha \beta} ), 
\end{eqnarray}
and $y_{J,K}(\alpha \alpha) = 0$ for odd $J$.
 Note the norm matrix is independent of $M$.
Now 
diagonalize
\begin{equation}
\sum_K \mathcal{N}^{(J)}_{K^\prime K} g^{(J)}_{K,r} = \nu^{(J)}_r  g^{(J)}_{K^\prime ,r}
\end{equation}
The number of nonzero eigenvalues $ \nu^{(J)}_r$  is the number of unique pairs.
Finally, we construct the unique collective pairs,
\begin{eqnarray}
\hat{A}^\dagger_{JM}(r) = 
( \nu^{(J)}_r)^{-1/2}  \sum_K g^{(J)}_{Kr}  \hat{B}^\dagger_{J,MK} \nonumber \\
= \sum_{\alpha \leq \beta} u^{(J)}_r(\alpha \beta)
\left [\hat{a}_{{\alpha}}^{\dagger} \times \hat{a}_{{\beta}}^{\dagger} \right]_{J,M}.
\end{eqnarray}
where 
\begin{equation}
u^{(J)}_r(\alpha \beta) = ( \nu^{(J)}_r)^{-1/2}  \sum_K g^{(J)}_{Kr}  y_{JK}(\alpha \beta).
\end{equation}
While these structure coefficients are calculated from a HF state with a particular
orientation, the final result is independent of that  orientation, which we confirmed numerically.


}

As mentioned above there is ambiguity in the choice of phases $g_{(2i-1)(2i)}$ in Eq. (\ref{HF}).
In this work, the phases are chosen so that the amplitude is maximized for $J = 0,2$.
In our calculations we have always found the amplitudes for $SD$ pairs to be large.


\section{Results and discussion}

\begin{figure}\center
\includegraphics[width = 0.24 \textwidth]{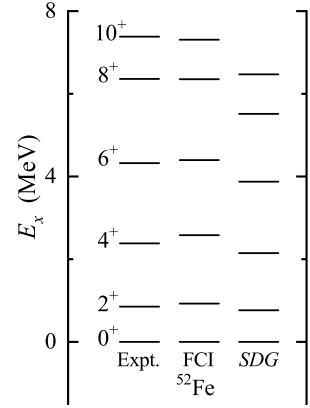}
\caption{\label{fig1}  Ground band of $^{52}$Fe.  Experimental data from \cite{a52nds}.
FCI = full configuration interaction, while $SDG$ is our NPA calculation.
}
\end{figure}
\begin{table}\center

 \begin{tabular}{|c|c|c|c|}  \hline\hline
    $I^{\pi}$  & Expt. & FCI &  $SDG$    \\  \hline
         $2^+$ & 14.2(19) & 16.0 & 12.9 \\
        $4^+$  & 26(6)    & 21.3 & 16.6 \\
        $6^+$  & 10(3)    & 11.8 & 13.4 \\
        $8^+$  & 9(4)     & 7.2  & 9.0 \\
        $10^+$ &          & 6.4  & 5.2 \\
       \hline\hline
 \end{tabular}
 \caption{\label{table1} $B(E2: I \rightarrow I-2)$ ( W.u.) for $^{52}$Fe  yrast states.
    }
 \end{table}

To test the validity of collective pairs derived from a HF state, we perform calculations for four rotational nuclei with valence nucleons outside doubly magic cores, both in the full configuration-interaction space (using the {\tt BIGSTICK } code \cite{BIGSTICK,johnson2018bigstick}) and in the NPA.
Specifically, we consider $^{52}$Fe in the $fp$ shell with the KB3G interaction \cite{kb3g} with a $^{40}$Ca core, $^{68}$Se and $^{68}$Ge in the $0f_{5/2}1p0g_{9/2}$ shell using the JUN45 interaction \cite{jun45} with a  $^{56}$Ni core, and $^{108}$Xe  in the $0g_{7/2}1d2s0h_{11/2}$ shell  with a $^{100}$Sn core using  the monopole-optimized effective interactions \cite{PhysRevC.86.044323,private1} based on the CD-Bonn potential renormalized by the perturbative $G$-matrix approach.
We also calculate the reduced electric quadrupole transition probability, for which we take the standard effective charges $(e_{\pi},e_{\nu})=(1.5, 0.5)$ for $^{52}$Fe and $^{108}$Xe, and $(1.5, 1.1)$ for $^{68}$Se and $^{68}$Ge.

Much of the motivation for the NPA is the significantly reduced dimensionality, and our purpose here is to validate this new approach for application of the NPA to heavy nuclei beyond the reach of FCI calculations.
Our FCI calculations are in the $M$-scheme (fixed total $J_z$), and the largest dimension calculation was $^{68}$Se, with an $J_z=0$ dimension of 165 million basis states.
{The largest NPA dimensionality, in fixed $J$-scheme, was the band-mixing (labeled $SDG$ (II) below) calculation for $^{68}$Se, for $J=4$, with a dimension of 7,253 basis states.}
{On a PC with an 8-core 4 GHz CPU, the $M$-scheme full configuration code {\tt BIGSTICK}  takes about 400 minutes,  while the $J$-scheme NPA code takes 4 minutes
in the truncated space.  Recent work \cite{he2020nucleons} suggests an $M$-scheme NPA code could run significantly faster, allowing one to reach much larger spaces.
For comparison, the $J$-scheme full-configuration code {\tt NuShellX } \cite{NuShellX} took about a day and a half (the untruncated $J=4$ space
has a dimension of 12.8 million basis states.) }


We start with $^{52}$Fe.
As discussed above, for an even-even nucleus the HF single-particle orbits come in degenerate time-reversed partners.
If the system is axially symmetric, those partners can have $z$-projection components $\pm m$, in which case the collective pair in the HF defined in Eq. (\ref{HF}) is restricted to $M=0$, and for each $J$ there is one unique pair.
From the prolate HF state of $^{52}$Fe  
 we extract one $S$ pair, one $D$ pair, and one $G$ pair.
The amplitude of $G$ pairs is non-negligible, and so in the NPA calculation of $^{52}$Fe, we construct our model space using
 $SDG$ pairs.

Fig.~\ref{fig1} and Table \ref{table1} compare for the  ground state band of $^{52}$Fe the experimental data \cite{a52nds}, the FCI, and the $SDG$-pair approximation results.
Both the level energies and the $B(E2)$ values obtained by the $SDG$ are in good agreement with the data or the FCI results, although the $SDG$ predicts a slightly larger {moment} of inertia and slightly smaller $BE2$ values for $2^+ \rightarrow 0^+$ and $4^+ \rightarrow 2^+$.

\begin{figure}\center
\includegraphics[width = 0.48 \textwidth]{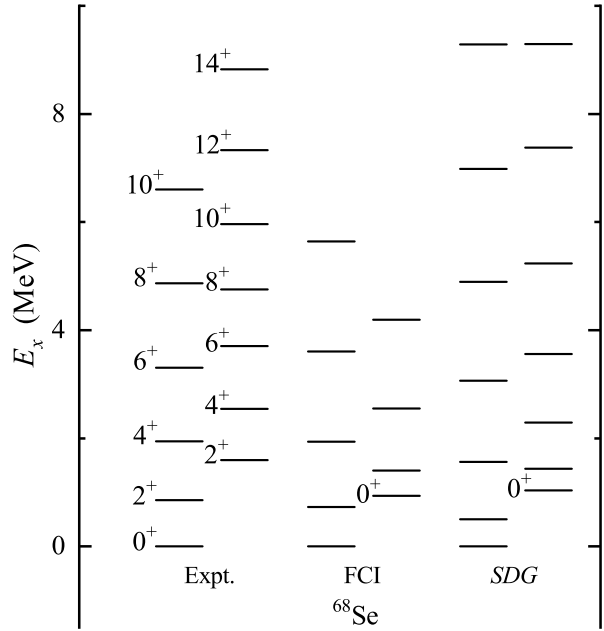}
\caption{\label{fig2} The ground rotational band (oblate) and the side band (prolate) of $^{68}$Se.
Experimental data from \cite{a68nds}.
}
\end{figure}

Shape coexistence has been experimentally observed in  $^{68}$Se  \cite{PhysRevLett.84.4064} and reproduced by the FCI calculation with the JUN45 interaction \cite{jun45}.
The ground rotational band is interpreted as an oblate deformation, and the low-lying side band as a prolate deformation.
{Our HF calculation produces an oblate minimum with $\langle \beta \rangle = 0.22$ and $\langle \gamma \rangle = 60^{\circ}$ and a  prolate one with $\langle \beta \rangle = 0.21$ and $\langle \gamma \rangle = 0^{\circ}$}, separated only by 900 keV.
{From the above HF states, we obtained two different sets of $SDG$ pairs, and the configuration spaces constructed by them are denoted by $\mathcal{L}_1$ and $\mathcal{L}_2$, respectively.
We carried out the NPA calculation of $^{68}$Se in two different ways:
(I) the oblate and prolate bands are calculated by diagonalizing the Hamiltonian in the $\mathcal{L}_1$ and $\mathcal{L}_2$ spaces, respectively;
(II) the oblate and prolate bands are calculated in the $\mathcal{L}_1 \bigoplus \mathcal{L}_2$ space, i.e., we mix the basis states from the two HF states.}

\begin{table}\center

 \begin{tabular}{|c|c|c|c|}  \hline\hline
    $I^{\pi}$  & Expt. & FCI &  $SDG$ (I)  \\  \hline 
 \multicolumn{4}{|c|}{ground (oblate) band} \\  \hline
         $2^+$ & 27(4) & 35.6 & 38.7  \\ 
        $4^+$  &       & 51.0 & 52.2  \\ 
        $6^+$  &       & 52.1 & 51.1  \\ 
        $8^+$  &       & 40.0 & 43.1 \\  \hline 
 \multicolumn{4}{|c|}{side (prolate) band} \\  \hline
         $2^+$ &       & 32.8 & 32.9  \\ 
        $4^+$  &       & 45.4 & 45.7  \\ 
        $6^+$  &       & 45.4 & 46.2  \\ 
        $8^+$  &       & 38.4 & 40.3  \\ 
       \hline\hline
 \end{tabular}
 \caption{\label{table2} $B(E2:I\rightarrow I-2)$ ( W.u.) for $^{68}$Se ground (oblate) and side (prolate) bands.
    }
 \end{table}

Fig. \ref{fig2} compares excitation energies from experiment \cite{a68nds}, the FCI calculations, and the $SDG$-pair approximation (I) and (II).
One sees that at low excitation energies the coexistence of the oblate and prolate bands is well reproduced by our $SDG$ pairs.
{The low-lying states calculated in $SDG$ (II) are very close to the $SDG$ (I) results, which means the configuration mixing between the oblate and prolate states is weak.}
Table \ref{table2} shows that the $B(E2)$ values in these two bands obtained by the $SDG$ (I) are very close to the FCI result.
The quadrupole moments of the $2^+_1$ and $2^+_2$ states calculated by the $SDG$ (I) are equal to $+51$ and $-48$ $e$-fm$^2$, which are also very close to the FCI result.
Both the FCI and the $SDG$ predict the prolate bandhead is a $0^+$ state, which has not yet been found experimentally.


\begin{figure}\center
	\includegraphics[width = 0.48 \textwidth]{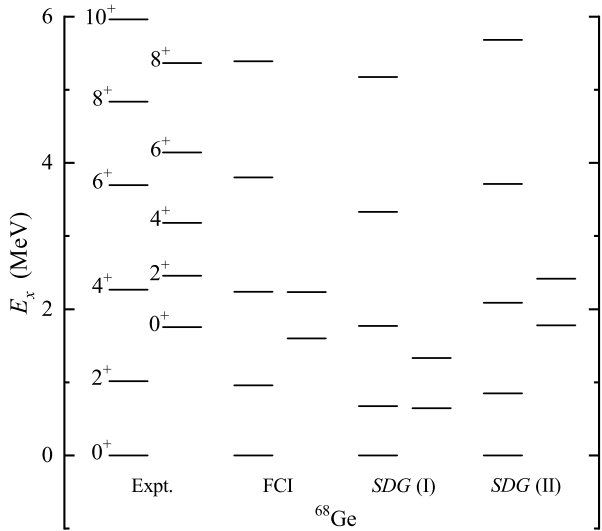}
	\caption{\label{fig2.2} The ground rotational band and the side band of $^{68}$Ge. Experimental data from \cite{a68nds}.
	}
\end{figure}

Similarly,  $^{68}$Ge also has a low-lying side band, starting with the $0^+_2$ state at 1.754 MeV (see Fig. \ref{fig2.2}).
{Our unconstrained, unrestricted HF calculation produces two minima differing in energy by only 1.114 MeV, both of which are triaxially deformed: the first minimum has $\langle \beta \rangle = 0.17$, $\langle \gamma \rangle = 38^\circ$, and the second one has $\langle \beta \rangle = 0.24$, $\langle \gamma \rangle = 44^\circ$.
We confirm these local minima are stable, as  the stability matrix, which is just the Tamm-Dancoff approximation matrix,
has only positive eigenvalues \cite{ring2004nuclear}.}
From the above HF states, we obtained two different sets of $SDG$ pairs, and the configuration spaces constructed by them are denoted by $\mathcal{L}_1$ and $\mathcal{L}_2$, respectively.
{Similar to the case of $^{68}$Se, the NPA calculation of $^{68}$Ge is carried out in two different ways, i.e., $SDG$-pair approximation (I) and (II).}

For $^{68}$Ge, Fig. \ref{fig2.2} and Table \ref{table2.2} compare experimental data \cite{a68nds}, the FCI, and the $SDG$ (I) and (II).
The low-lying states calculated in $SDG$ (II) are in good agreement with the FCI results, but those from $SDG$ (I) are not.
For example, the excitation energy of the $0^+_2$ state in  $SDG$ (II) is 1.781 MeV,   close to experimental data, but that from $SDG$ (I) is only 0.644 MeV.
The $B(E2)$ values given by $SDG$ (II) are very close to the FCI results, but $SDG$ (I) yields smaller values for the ground band and a larger value for the side band.
The above results indicate that the configuration mixing between the two different HF states is important in the ground and side bands of $^{68}$Ge.

\begin{figure}\center
\includegraphics[width = 0.40 \textwidth]{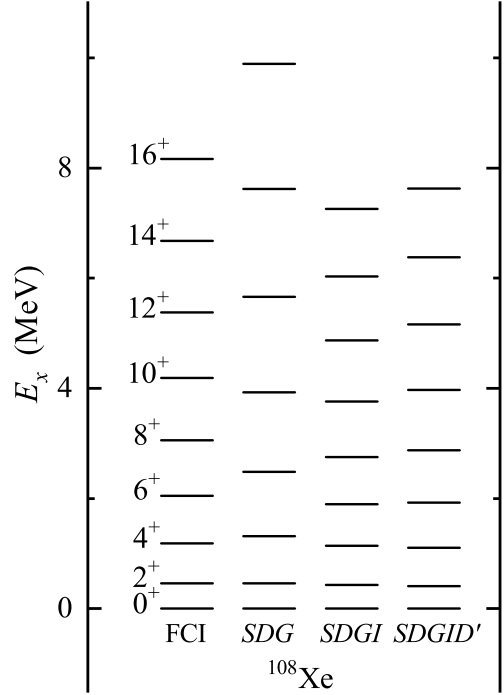}
\caption{\label{fig3} The ground rotational band and $B(E2:I\rightarrow I-2)$ of $^{108}$Xe.
}
\end{figure}

The $N = Z$ isotope of $^{108}$Xe has been observed recently \cite{PhysRevLett.121.182501}, but the low-lying spectrum has not yet been experimentally studied.
Our HF calculation  of $^{108}$Xe has  triaxial deformation ($\langle \gamma \rangle = 11^{\circ}$).
From this HF state, we obtain one $S$ pair, two $D$ pairs, two $G$ pairs, and two $I$ pairs (collective pairs with spin six).
The amplitudes of the second $DGI$ pairs are relatively much smaller than those of the first ones.
For $^{108}$Xe, we focused on the ground rotational band, and thus our NPA  model space is constructed from only the first $SDG$ pairs.
Since the amplitude of the first $I$ pair is non-negligible, we also perform an NPA calculation in the space constructed by using the first $SDGI$ pairs.

Fig.~\ref{fig3} compares the excitation energies and $B(E2)$ values between the FCI results, the $SDG$- and $SDGI$-pair-approximation results for $^{108}$Xe.
The level energies of the low-lying $2^+$ and $4^+$ states obtained by the $SDG$ are in quite good agreement with the FCI results, and the same to the $B(E2)$ values for $2^+ \rightarrow 0^+$ and $4^+ \rightarrow 2^+$.
However,  for higher-spin states we see increasing discrepancy, suggesting
the collective $I$ pair may be important.
Indeed, for level energies and $B(E2)$ values, the agreement between the $SDGI$ and the FCI results are significantly improved, even if the former predict a moment of inertia slightly larger than the latter { and $B(E2)$ values slightly smaller than the latter.}
While these results are satisfactory, if the second $D$ pair, which appears because of triaxial deformation, is included in the basis states (for simplicity the maximum number of the second $D$ pair is constrained to one), results are further improved (see $SDGID^{\prime}$ in Fig. \ref{fig3}).





\begin{table}\center
	\begin{tabular}{|c|c|c|c|c|}  \hline\hline
		$I^{\pi}$  & Expt. & FCI &  $SDG$ (I)  &  $SDG$ (II)    \\  \hline
		\multicolumn{5}{|c|}{ground band} \\  \hline
		$2^+$ &15.3(8) & 28.1 & 23.9 & 25.4 \\
		$4^+$  &12.8(15)& 38.6 & 31.3 & 35.1 \\
		$6^+$  & 12(4)  & 44.9 & 28.4 & 35.2 \\
		$8^+$  & 14(3)  & 32.6 & 13.8 & 28.6 \\  \hline
		\multicolumn{5}{|c|}{side band} \\  \hline
		$2^+$ & 22(7) & 21.8 & 35.3 & 24.1 \\
		\hline\hline
	\end{tabular}
	\caption{\label{table2.2} $B(E2:I\rightarrow I-2)$ (W.u.) for $^{68}$Ge  ground and side bands. In $SDG$ (I) the bands
		are computed separately; in $SDG$ (II) they are mixed.
	}
\end{table}

\section{Summary and acknowledgements}

In this paper, we propose a simple and practical approach to generate collective nucleon pairs of good angular momentum for realistic NPA calculations for even-even rotational nuclei.
We recast HF states, computed in a shell-model basis, as a pair condensate, from which we project out pairs of
good angular momentum.
Applying this method to calculations of  $^{52}$Fe, $^{68}$Se, $^{68}$Ge, and $^{108}$Xe with effective interactions,
we find that  the $SDG$ pairs obtained by our approach provide us with good descriptions for low-lying states of the rotational bands and the phenomenon of shape coexistence, and that a high-spin $I$ pair is responsible for high-spin states of $^{108}$Xe.

One can generalize this approach further.
For example, if one replaces the pair condensate in Eq.~(\ref{HF}) with a wave function
\begin{eqnarray}
 \left( \hat{c}_{1}^{\dagger} \hat{c}_{2}^{\dagger} + \cdots + \hat{c}_{2\Omega-1}^{\dagger} \hat{c}_{2\Omega}^{\dagger} \right)^{N} |0\rangle,
\end{eqnarray}
where $2\Omega$ is the number of single-particle states in the space, one has something akin to a seniority-zero wave function.
One can also replace Eq. (\ref{HF}) with a number-projected BCS wave function
\begin{eqnarray}
 \left(  \sum_{  a} g_{ a\bar{a}} ~ \hat{c}^{\dagger}_{a} \hat{c}^{\dagger}_{\bar{a}} \right)^{N} |0\rangle,
\end{eqnarray}
where $a\bar{a}$ are time-reversed orbits, and $g_{a\bar{a}}$ is the occupation probability.
The generalization with the number-projected BCS is reasonably expected to further improve validity of the NPA.

It should be noted that for rotational nuclei, the NPA truncates the shell model configuration space in the spherical single-particle basis, while the adopted collective pairs are projected out from the deformed HF, connecting the spherical shell model with deformed models.
This work also suggests the microscopic foundation of the interacting boson model for deformed nuclei in terms of nucleon degree of freedom.
A boson mapping from shell model effective interactions would be very interesting.

With this approach the NPA can be a practical and powerful truncation scheme of the shell model to study quadrupole deformation, nuclear shape-phase transition, and octupole collectivity in low-lying states of heavy nuclei which are difficult to be realized in the large-scale FCI due to huge dimensions of configuration space.



This material is based upon work supported by the U.S. Department of Energy, Office of Science, Office of Nuclear Physics, under Award No. DE-FG02-03ER41272, the National Natural Science Foundation of China under Grant No. 11605122, and the National Key R\&D Program of China under Grant No. 2018YFA0404403. This collaboration was initiated through CUSTIPEN (China-U.S. Theory Institute for Physics with Exotic Nuclei) funded by the U.S. Department of Energy, Office of Science grant number DE-SC0009971.

\bibliographystyle{apsrev4-1}
\bibliography{johnsonmaster}

\end{document}